\documentclass[aps,pra,twocolumn,groupedaddress,amsmath,amssymb]{revtex4-1}

\usepackage[colorlinks,pdfusetitle,urlcolor=blue,citecolor=blue,linkcolor=blue,bookmarksnumbered,plainpages=false]{hyperref}

\usepackage{color}

\usepackage{bm}
\usepackage{nicefrac}
\usepackage{siunitx}

\usepackage{graphicx}
\usepackage{bm}
\usepackage{braket}

\newcommand{\B}[1]{\mathbf{#1}}

\newcommand{\sinAbbrv}{\eta}
\newcommand{\cosAbbrv}{\chi}

\newcommand{\wholeGF}{ \B{W}}

\begin{document}

\title{Spontaneous decay rate and Casimir-Polder potential of an atom near a lithographed surface}

\author{Robert Bennett}

\affiliation{Department of Physics \& Astronomy, University of Leeds, LS2 9JT, UK}

\date{\today}

\begin{abstract} Radiative corrections to an atom are calculated near a half-space that has arbitrarily-shaped small depositions upon its surface. The method is based on calculation of the classical Green's function of the macroscopic Maxwell equations near an arbitrarily perturbed half-space using a Born series expansion about the bare half-space Green's function. The formalism of macroscopic quantum electrodynamics is used to carry this over into the quantum picture. The broad utility of the calculated Green's function is demonstrated by using it to calculate two quantities --- the spontaneous decay rate of an atom near a sharp surface feature, and the Casimir-Polder potential of a finite grating deposited on a substrate. Qualitatively new behaviour is found in both cases, most notably in the latter where it is observed that the periodicity of the Casimir-Polder potential persists even outside the immediate vicinity of the grating.    \end{abstract}

\pacs{}

\maketitle

\section{Introduction}

Quantum fluctuations of the electromagnetic field are influenced by material boundaries, meaning that a wide variety of quantum electrodynamical vacuum effects have an environment-dependence which are often referred to as dispersion forces.  Famous examples include the force between macroscopic objects known as the Casimir effect \cite{casimir1948attraction}, and the closely-related Casimir-Polder force \cite{Casimir:1948bd} between an atom and a surface. Other examples include modified spontaneous decay rates \cite{Yeung:1996fp, Scheel:1999eq}, magnetic moments \cite{Bennett:2012va, Bennett:2013is}, cyclotron frequencies \cite{Bennett:2012ca, Barton:1988tz} and Zeeman splittings \cite{Bennett:2014fs, Donaire:2014ws}. 

There is contemporary interest in how dispersion forces are modified by the specifics of the surfaces involved. This can be by consideration of their optical properties \cite{Rosa:2008ks}, their thermal environment \cite{Antezza:2005dl} or their geometries. An example of the latter is found in \cite{Dalvit:2008jt} where it is shown that nontrivial geometry-dependent vacuum effects can be studied by using a Bose-Einstein condensate above a corrugated surface. Dispersion-force calculations that go beyond simple planar geometries are usually complicated in the extreme due to the inherent non-additivity of dispersion forces (see, for example, \cite{milonni1994quantum, Farina:1999kr}). To remedy this, various simplifying approaches have been developed, one of the most prominent being the `proximity-force approximation' (PFA) \cite{derjaguin1957direct} where one models complex geometries as made up of an ensemble of flat, parallel surfaces. It has been numerously shown (see, for example, \cite{ContrerasReyes:2010bz, Gies:2006di, Rodriguez:2007ul, Neto:2007gk, Reynaud:2008ff}) that the PFA is uncontrolled and is often significantly in error. There are other approaches based on surface being `almost smooth' \cite{Bimonte:2014fw, Messina:2009gv}, but none are readily applicable to mechanically-etched surfaces with sharp edges like those discussed in  \cite{Nshii:2013fv}, for example. 

Here we will use an alternative method based on the Born expansion of the Green's function of the electromagnetic wave equation, which will be used to calculate environment-modified decay rates and Casimir-Polder potentials near a selection of geometries. In contrast to the PFA, this approach preserves the rich geometry-dependence of dispersion forces, at the expense of requiring the system to consist of a small `geometric perturbation' from an exactly solvable `background' geometry.  This approach has been used before in the calculation of Casimir-Polder potentials \cite{Buhmann:2005ioa} and Casimir forces \cite{Golestanian:2009gv, Bennett:2014gs}. One of the main differences between our work and \cite{Buhmann:2005ioa,Golestanian:2009gv,Bennett:2014gs} is that only homogenous backgrounds were considered there, while we consider a half-space as the background. The advantage of this is that the optical properties of the half-space can be specified completely freely --- it is not part of the perturbation so its electromagnetic response does not need to satisfy any of the conditions that ensure convergence of the perturbation series.  This allows one to make perturbative calculations for quantum electrodynamical quantities near arbitrarily-shaped small depositions onto the surface of the (non-perturbative) half-space, which is the goal of this paper.  These kinds of geometries are relevant to very recent experiments on decay rates near pattered materials \cite{Lu:2014fa}, and could also be applied to studies of surface roughness \cite{Suresh:1996kj, Bezerra:2000fu}. 

\section{Theoretical Background}

We will use the noise-current approach \cite{Gruner:1996cw, Dung:1998ic} to electromagnetic (EM) field quantisation in and around dielectric media. This approach is necessitated by the fact that Maxwell's equations in a dispersive, absorbing medium cannot be quantised simply by promoting the field observables to operators as this would cause a violation of the fluctuation-dissipation theorem. To remedy this, one introduces a source current density operator $\B{j}$ which corresponds to noise associated with loss in the medium and restores consistency with the fluctuation-dissipation theorem \cite{Matloob:1995fm,Gruner:1996cw, Dung:1998ic,Scheel:2008tu}. It is interesting to note that in its original form this theory did not rest on a rigorous canonical foundation, however this was recently remedied in \cite{Philbin:2010fa}. The advantage of the use of this source current representation is that it allows the quantised field to be obtained from the classical Green's function for the electromagnetic field in a given configuration  \cite{Matloob:1995fm, Gruner:1996cw, Dung:1998ic,Scheel:2008tu}. In this framework, the frequency-domain \emph{quantised} electric field that solves Maxwell's equations in a medium with position and frequency-dependent permittivity $\epsilon(\B{r},\omega)$ is given by the solution to the following wave equation \footnote{We work in a system of natural units where the speed of light $c$, the reduced Planck constant $\hbar$ and the permittivity of free space $\epsilon_0$ are all equal to $1$. }
 \begin{equation} \label{BasicWaveEqn}
 \nabla \times \nabla \times \B{E}(\B{r},\omega) - \omega^2 \epsilon(\B{r},\omega) \B{E}(\B{r},\omega) = i \omega \B{j} (\B{r},\omega) \, , 
\end{equation}
with $\B{j}$ being the operator-valued noise-current source discussed above. This can be solved by the introduction of a Green's function (variously called the dyadic Green's function, or the Green's tensor) \cite{Gruner:1996cw, Dung:1998ic} which we will call $\wholeGF(\B{r},\B{r}',\omega)$ \footnote{The reason for avoiding the standard notation $\B{G}$ is that we reserve this symbol for the scattering Green's function (consistent with our previous work \cite{Bennett:2014gs}), as opposed to the whole Green's function. In \cite{Bennett:2014gs} the whole Green's function $\wholeGF$ was given the more obvious symbol $\B{\Gamma}$, but here that is reserved for the spontaneous decay rate.}. It is defined as the solution to
\begin{align}
\nabla \times \nabla \times \wholeGF(&\B{r},\B{r}',\omega) \notag \\
&- \omega^2 \epsilon(\B{r},\omega)  \wholeGF(\B{r},\B{r}',\omega) = \mathbb{I} \delta(\B{r}-\B{r}') \, , \label{GFDefiningEq}
\end{align}
where $\mathbb{I}$ is a $3\times 3$ unit matrix. 

The Green's function $\wholeGF$ defined by \eqref{GFDefiningEq} uniquely determines the quantised field in a particular configuration, which ultimately means that knowledge of $\wholeGF$ allows one to calculate a wide variety of quantum electrodynamical quantities. However, exact calculation of the Green's function $\wholeGF$ is only possible analytically for the very simplest choices of $\epsilon(\B{r},\omega)$, so here we avoid this problem by using a perturbative technique. As shown in \cite{Buhmann:2005ioa} one can write the unknown $\wholeGF$ in terms of some known `background' Green's function $\wholeGF^{(0)}(\B{r},\B{r}',\omega) $ as
\begin{align}\label{BornSeriesNotSimp}
\wholeGF(& \B{r},  \B{r}',\omega) = \wholeGF^{(0)}(\B{r},\B{r}',\omega) \notag\\ &+\omega^2 \int d^3 \B{s}_1 \wholeGF^{(0)}(\B{r},\B{s}_1,\omega) \delta \epsilon(\B{s}_1,\omega)  \wholeGF^{(0)}(\B{s}_1,\B{r}',\omega)\notag \\
&+\omega^4 \int d^3 \B{s}_1\int d^3 \B{s}_2 \Big[ \wholeGF^{(0)}(\B{r},\B{s}_1,\omega) \delta \epsilon(\B{s}_1,\omega)\notag\\
& \times \wholeGF^{(0)}(\B{s}_1,\B{s}_2,\omega) \delta \epsilon(\B{s}_2,\omega)  \wholeGF^{(0)}(\B{s}_2,\B{r}',\omega)\Big]+... \; , 
\end{align}
where $\delta \epsilon(\B{r}, \omega)$ is the difference between the entire dielectric function and that of the background material at a particular point $\B{r}$. This type of perturbative expansion is known as the Born series and is the foundation of much of scattering theory --- the spatial integrations over $\B{s}_i$ have a definite interpretation as scattering events \cite{Buhmann:2005ioa, Golestanian:2009gv, Scheel:2008tu, Bennett:2014gs}. 

\begin{figure}[h!]
\includegraphics[width = 0.5\columnwidth]{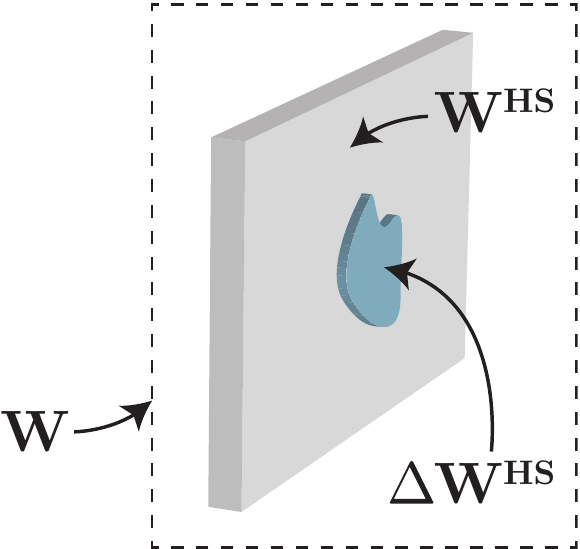}
\caption{General setup} \label{GeneralPerturbationDiagram}
\end{figure}

We can simplify the Born series \eqref{BornSeriesNotSimp} by specifying that the configurations we are interested in are always made up of an object described by some volume $\B{V}$ that has an internally homogenous dielectric function $ \epsilon(\B{r},\omega) = \epsilon(\omega)$, and sits in some (possibly inhomogenous) `background'  material with dielectric function $\epsilon^{(0)}(\B{r},\omega)$. Under these assumptions we can restrict the $\B{s}_i$ integrals to being over the volume $\B{V}$, because outside this region the background dielectric function at a particular point is equal to the entire dielectric function at that point, so $\delta \epsilon(\B{r},\omega)=0$ there. The assumption of homogeneity within the volume $\B{V}$ means we can also bring the dielectric functions outside the integrals, giving
\begin{align}
&\wholeGF(\B{r},  \B{r}',\omega) = \wholeGF^{(0)}(\B{r},\B{r}',\omega) \notag\\ 
&+\omega^2 [\delta \epsilon(\omega)]  \int_\B{V} d^3 \B{s}_1 \wholeGF^{(0)}(\B{r},\B{s}_1,\omega)  \wholeGF^{(0)}(\B{s}_1,\B{r}',\omega)\notag \\
&+\omega^4 [\delta \epsilon(\omega) ] ^2 \int_\B{V} d^3 \B{s}_1\int_\B{V} d^3 \B{s}_2 \Big[ \wholeGF^{(0)}(\B{r},\B{s}_1,\omega) \notag\\
& \times \wholeGF^{(0)}(\B{s}_1,\B{s}_2,\omega)   \wholeGF^{(0)}(\B{s}_2,\B{r}',\omega)\Big]+... \, . 
\label{ScatteringDerivation} \end{align}
%
%
In order to work out surface-modified quantities we will need the so-called `scattering' part of the Green's function \eqref{ScatteringDerivation} --- that is, the part which remains after the subtraction of the Green's function for a homogenous region. We write the scattering part of the \emph{whole} Green's function $\wholeGF$ as $\B{G}$ and the remaining \emph{homogenous} part as $\B{H}$. This means the whole Green's function can be rewritten
\begin{align}
&\wholeGF(\B{r},  \B{r}',\omega) = \B{G}^{(0)}(\B{r},\B{r}',\omega)+\B{H}^{(0)}(\B{r},\B{r}',\omega) \notag\\ 
&+\omega^2 [\delta \epsilon(\omega)]  \int_\B{V} d^3 \B{s}_1 \left[ \B{G}^{(0)}(\B{r},\B{s},\omega)+\B{H}^{(0)}(\B{r},\B{s},\omega)\right]\notag \\
&\qquad \qquad \cdot \left[  \B{G}^{(0)}(\B{s},\B{r}',\omega)+\B{H}^{(0)}(\B{s},\B{r}',\omega) \right]+... \; . 
 \end{align}
 In previous calculations \cite{Buhmann:2005ioa,Golestanian:2009gv,Bennett:2014gs} the background Green's function $\wholeGF^{(0)}$ was taken to be that for a homogenous medium, so that its scattering part $\B{G}^{(0)}$ is by definition zero. This has the simplifying property that the partitioning of the Green's function via the Born series coincides with the partitioning one makes when finding the scattering part, i.e. for homogenous $\B{H}$
 \begin{align}
&\wholeGF_{\B{H}\, \text{hom}}(\B{r},  \B{r}',\omega) = \B{H}^{(0)}(\B{r},\B{r}',\omega) \notag\\ 
&\qquad +\omega^2 [\delta \epsilon(\omega)]  \int_\B{V} d^3 \B{s}_1 \B{H}^{(0)}(\B{r},\B{s},\omega)\B{H}^{(0)}(\B{s},\B{r}',\omega) \, ,
 \end{align}
 meaning that the scattering part is:
  \begin{align}
\B{G}_{\B{H}\, \text{hom}}&(\B{r},  \B{r}',\omega)= \notag \\
&\omega^2 [\delta \epsilon(\omega)]  \int_\B{V} d^3 \B{s}_1 \B{H}^{(0)}(\B{r},\B{s},\omega)\B{H}^{(0)}(\B{s},\B{r}',\omega) \, . 
 \end{align}
 This means that all that is required for the calculation of environment-dependent quantities in a geometry regarded as a perturbation to a \emph{homogenous} medium is an integral over a homogenous Green's function, which is relatively simple to do. However this is not usually a case of physical interest since in real experiments there will likely be an object nearby that does not obey the conditions for convergence that the Born expansion requires. To remedy this, we will study the simplest \emph{inhomogenous} background, namely a half-space, and then add perturbing objects to \emph{that}, as shown in Fig.~\ref{GeneralPerturbationDiagram}. 

In this work we will truncate the Born series at the single-scattering term, though it is straightforward to extend the method to higher-order terms as is required to work out Casimir forces \cite{Bennett:2014gs} as opposed to decay rates and Casimir-Polder potentials as is done here. We then have the whole Green's function to order $\delta \epsilon(\B{r},\omega)$
\begin{align}
\wholeGF(\B{r},  \B{r}',\omega) &= \wholeGF^{\text{HS}}(\B{r},\B{r}',\omega)\notag\\
+\omega^2& [\delta \epsilon(\omega)]  \int_V d^3 \B{s} \wholeGF^{\text{HS}}(\B{r},\B{s},\omega)   \wholeGF^{\text{HS}}(\B{s},\B{r}',\omega)\notag \\
&=\wholeGF^{\text{HS}}(\B{r},\B{r}',\omega)+\Delta\wholeGF^{\text{HS}}(\B{r},\B{r}',\omega) \; ,  \label{BornExp} \end{align}
with $ \wholeGF^{\text{HS}}$ being the Green's function for a half-space. The half-space Green's function at frequency $\omega$ in a region $z>0$ in the presence of a non-magnetic material half-space filling the region $z<0$ is conveniently written as \cite{ChewDGFBook}:
\begin{align} \label{HalfSpaceGF}
&\wholeGF^{\text{HS}}(\B{r},  \B{r}',\omega) =-\frac{\hat{z} \otimes \hat{z}}{k^2} \delta(\B{r}-\B{r}') \notag \\
 &+ \frac{i}{8\pi^2} \sum_{\sigma} \int d^2 \B{k}_\parallel \frac{ \mathcal{D}_\sigma (\B{r},\B{r'})}{k_\parallel^2 k_{z}}e^{i\B{k}_\parallel \cdot (\B{r}_\parallel - \B{r}'_\parallel )}F_\pm^\sigma(z,z') \, , \end{align}
where $k_z \equiv \sqrt{\omega^2-k_\parallel^2}$ and $\B{r}_\parallel$ and $\B{k}_\parallel$ are respectively the components  of the position and wave vector parallel and perpendicular to the interface, and $\hat{z}$ is a unit vector perpendicular to the interface. The symbol $\sigma$ indexes the two possible polarisations [TE (transverse-electric) and TM (transverse-magnetic)] of the Coulomb-gauge electromagnetic field, and  $\mathcal{D}_\sigma$ represents the following differential operators
\begin{align} \label{DOps}
\mathcal{D}_\text{TE}(\B{r},\B{r}') &\equiv \left(\nabla \times \hat{z}\right) \otimes \left(\nabla' \times \hat{z}\right) \, , \notag  \\
\mathcal{D}_\text{TM}(\B{r},\B{r}') &\equiv \frac{1}{\omega^2} \left(\nabla \times \nabla \times \hat{z}\right) \otimes \left(\nabla'\times \nabla' \times \hat{z}\right)  \, . 
\end{align}
Finally, the function $F_\pm^\sigma(z,z')$ is given by
\begin{align}
F^\sigma(z,z') &= \left[ e^{- ik_{z} \mathcal{Z}_<}+e^{ ik_{z}\mathcal{Z}_<} R^\sigma_{vm}\right]  e^{ ik_{z} \mathcal{Z}_>} \, ,  \label{FDefs} \end{align}
where $\mathcal{Z}_>$ is the greater of $z$ and $z'$, and $\mathcal{Z}_<$ is the lesser of $z$ and $z'$;
\begin{align}
\mathcal{Z}_> &= \begin{cases} z &\text{ for } z>z'\\
z' &\text{ for } z<z'
\end{cases}, &  \mathcal{Z}_< &= \begin{cases} z' &\text{ for } z>z'\\
z &\text{ for } z<z'
\end{cases} \; ,
\end{align}
and $R^\sigma_{vs}$ are the Fresnel coefficients for radiation propagating from a vacuum region into a medium of permittivity  $\epsilon(\omega)$
\begin{align}
R_{vm}^\text{TE} &= \frac{k_z-k_z^d}{k_z+k_z^d}, & R_{vm}^\text{TM} &= \frac{\epsilon(\omega) k_z-k_z^d}{\epsilon(\omega) k_z+k_z^d} ,
\end{align}
where $k_z^d=\sqrt{\epsilon(\omega)\omega ^2-k_\parallel^2}$ is the $z$-component of the wave vector inside the medium. We can now use this statement of the half-space Green's function to generate the next-to-leading order term $\Delta\wholeGF^{\text{HS}}(\B{r},\B{r}',\omega) $ in the Born expansion \eqref{BornExp}, which will give the modified Green's function for the EM field in the vicinity of a half-space with depositions.

\section{Modified Green's function}

We now  present the Green's function modification $\Delta\wholeGF^{\text{HS}}(\B{r},\B{r}',\omega)$ for a half-space with a deposition. We will restrict ourselves to the region $\B{r},\B{r}' \neq \B{s}$ throughout this work, meaning that we can ignore the $\delta$ function terms in \eqref{HalfSpaceGF} when substituting it into \eqref{BornExp}. This means that we will not calculate any quantum electrodynamical quantities \emph{inside} a deposition onto a half-space. Apart from complicating the method used here, calculation of such quantities would require the use of local-field corrected Green's tensors \cite{Knoester:1989kk, Barnett:1992gf, Scheel:1999eq} which are beyond the scope of this work. Under these assumptions, we note that $\Delta \wholeGF(\B{r},\B{r}',\omega)$ depends quadratically on $F^\sigma$, so from the form of Eq.~\eqref{FDefs} one sees that that all  contributions to $\Delta\wholeGF$ as defined in Eq.~\eqref{BornExp} must be at most quadratic in the reflection coefficients, so we can write
\begin{align} \label{ModGFunction}
&\Delta\wholeGF^{\text{HS}}_{ij}(\B{r},\B{r}',\omega) =  \int_\B{V} d^3 \B{s}  \int d^2  \B{k}_\parallel  \int d^2 \B{k}'_\parallel P  \notag \\
&\!\!\!\!\times \Big[1\!+\!K_{ij\lessgtr}^\text{TETM} R_\text{TE} R_\text{TM}\!+\!\sum_{\sigma} (K^\sigma_{ij\lessgtr} R_\sigma\!+\!K^{\sigma\sigma}_{ij\lessgtr} R^2_\sigma) \Big],\end{align}
where, for later convenience, we have  defined the quantity $P$  as 
 \begin{align}
&P= -\frac{ \delta \epsilon(\omega) }{64\pi^4 { \omega ^4 k_\parallel^2 k_\parallel^{'2}  k_zk_z'}}\exp \Big\{i \Big[\B{k}_\parallel \cdot (\B{r}_\parallel-\B{s}_\parallel) \notag \\
&+ \B{k}'_\parallel \cdot  (\B{s}_\parallel-\B{r}'_\parallel) + k_z(r_z+s_z)  + k_z'(r'_z + s_z) 
\Big]\Big\}\, . 
\end{align}
The various $K_{ij\lessgtr}$ in \eqref{ModGFunction} are matrix elements determined from Eqs.~\eqref{BornExp} and \eqref{HalfSpaceGF} by simple but tedious application of the differential operators \eqref{DOps} to the functions $F_\pm^\sigma(z,z')$ defined in Eq.~\eqref{FDefs}. The matrix elements differ depending on wether $r_z$ is greater or less than $s_z$, the subscript $\lessgtr$ distinguishes these two cases, as detailed in the full list of matrix elements found in Appendix \ref{GFMatrixElements}.

\subsection{Simple demonstration: Decay rate near a sharp surface feature}\label{DecayRateSection}

We will begin with a surface-modified quantity that is relatively easy to calculate, namely the spontaneous decay rate $\Gamma$ of an excited atom that is attributable to its interaction with the quantised electromagnetic field. It has been shown \cite{Scheel:1999js} that this rate can be expressed in terms of the Green's function $\wholeGF$ as:
\begin{equation} \label{DecayRateBasicExpression}
\Gamma =2\omega_\text{A}^2 \B{d} \cdot \left[ \text{Im} \wholeGF(\B{r}_\text{A},\B{r}_\text{A},\omega_\text{A})\right] \cdot \B{d}^* \, , 
\end{equation}
where $\B{d}$ is the dipole moment of the transition. As an example we will calculate the decay rate $\Gamma_0$ of an atom in vacuum, with no material objects present. We can choose the direction of the polarisation freely because in vacuum we have rotation invariance --- we choose the polarisation to be aligned along the $\B{z}$ direction so that $\B{d} = d \hat{\B{z}}$. Then;
\begin{equation}
\Gamma_0 = 2\omega_\text{A}^2 |d|^2  \text{Im} \wholeGF^\text{vac}_{zz}(\B{r}_\text{A},\B{r}_\text{A},\omega_\text{A}) \, ,  \label{FreeSpaceDecayRate}
\end{equation}
where $\wholeGF_{zz}^\text{vac}$ is the $zz$ component of the Green's function that solves \eqref{GFDefiningEq} for $\epsilon(\B{r},\omega) = 1$. This vacuum Green's function is well-known (see \cite{Scheel:2008tu} for a thorough review). It can be found, for example, from the half-space Green's function \eqref{HalfSpaceGF} reported here by taking all reflection coefficients to zero;
\begin{equation}
\wholeGF^{\text{vac}}(\B{r},\B{r}',\omega) = \wholeGF^\text{HS}(\B{r},\B{r}',\omega)|_{R^\sigma = 0} \; . 
\end{equation}
The $zz$ component of the vacuum Green's function is
\begin{equation}
\wholeGF^\text{vac}_{zz}(\B{r},\B{r}',\omega_\text{A})=\frac{i}{8 \pi ^2}\int d^2 \B{k_\parallel} \frac{k_\parallel^2}{  \omega ^2 k_z} e^{ik_z|z-z'|} \; , 
\end{equation}
where we have ignored the (real-valued) $\delta$ function part of \eqref{HalfSpaceGF} in anticipation of taking the imaginary part as dictated by \eqref{DecayRateBasicExpression}. Transforming to polar co-ordinates in the $k_x,k_y$ plane and doing the trivial angular integral we have
\begin{equation}
\wholeGF^\text{vac}_{zz}(\B{r},\B{r}',\omega_\text{A})=\frac{i}{4 \pi}\int_0^\infty d {k_\parallel} \frac{k_\parallel^3}{  \omega ^2 k_z} e^{ik_z|z-z'|} \; . 
\end{equation}
The integral can be carried out analytically, giving:
\begin{equation}
\wholeGF^\text{vac}_{zz}(\B{r},\B{r}',\omega_\text{A}) = \frac{1}{2\pi \omega^2} \frac{e^{i \omega  |z-z'|}}{ |z-z'|^3} (1-i \omega  |z-z'|) \; . 
\end{equation}
Taking the imaginary part of this followed by the coincidence limit $z' \to z$, we find upon substitution into \eqref{FreeSpaceDecayRate}
\begin{equation}
\Gamma_0 = \frac{\omega_\text{A}^3}{3 \pi } |d|^2 \; , 
\end{equation}
which is a well-known result (see, for example, \cite{vogel2006quantum}), and will be used as a convenient unit in the following discussions.

\begin{figure}[h!]
\includegraphics[width = \columnwidth]{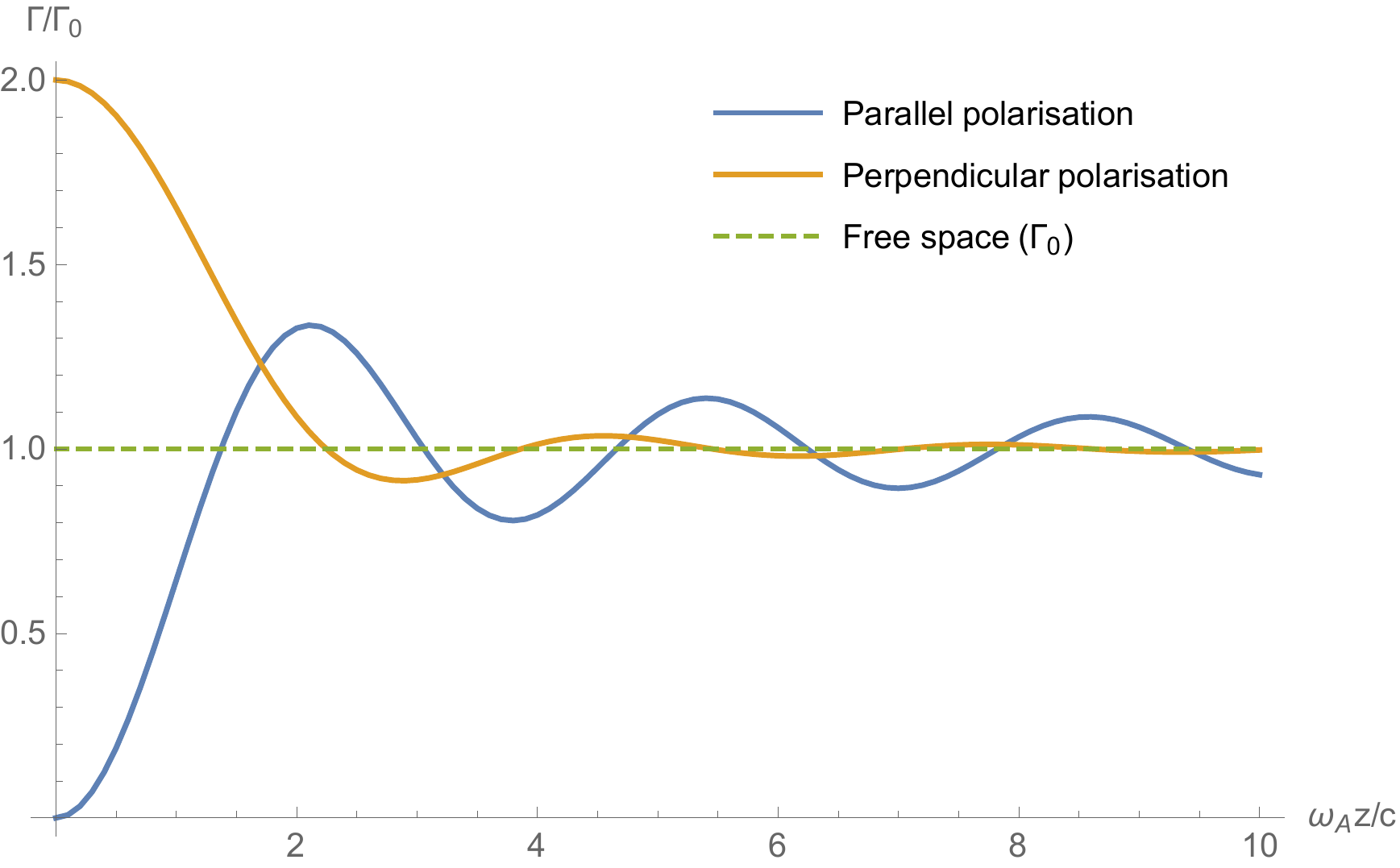}
\caption{Decay rates near a simple half-space} \label{DecayRateHalfSpace}
\end{figure}

As another point of comparison we also present the results for the decay rate near a half-space
\begin{equation}
\Gamma^\text{HS} = 2\omega_\text{A}^2\B{d} \cdot \left[ \text{Im} \wholeGF^\text{HS}(\B{r}_\text{A},\B{r}_\text{A},\omega_\text{A})\right] \cdot \B{d}^*\notag \; , 
\end{equation}
which we split into the free-space contribution $ \Gamma_0$ and a surface-modified part $\Delta \Gamma^{\text{HS}}$: 
\begin{align}
\Gamma^\text{HS} &= 2\omega_\text{A}^2 \B{d} \cdot \Big[ \text{Im}( \wholeGF_{\text{vac}}(\B{r}_\text{A},\B{r}_\text{A},\omega_\text{A})\notag \\
&\qquad \qquad \qquad \qquad+\B{G}^\text{HS}(\B{r}_\text{A},\B{r}_\text{A},\omega_\text{A})\Big] \cdot \B{d}^*\notag \\
&= \frac{\omega_\text{A}^3|d^2|}{3 \pi }+ {2\omega_\text{A}^2} \B{d}\cdot  \text{Im}\B{G}^\text{HS}(\B{r}_\text{A},\B{r}_\text{A},\omega_\text{A})\cdot \B{d}^* \notag \\
& = \Gamma_0 + \Delta \Gamma^{\text{HS}} \; . 
\end{align}
We will consider the two cases that the atom is polarized parallel and perpendicular to the surface with the same magnitude of dipole moment $d = |\B{d}|$, and separately find the two contributions $ \Delta \Gamma^{\text{HS}}_\parallel$ and $ \Delta \Gamma^{\text{HS}}_\perp$ to the decay rates $\Gamma_\parallel = \Gamma_0 + \Delta \Gamma^{\text{HS}}_\parallel $ and $\Gamma_\perp = \Gamma_0 + \Delta \Gamma^{\text{HS}}_\perp $. Calculation of $\Delta \Gamma_{\parallel}^\text{HS}$ is simplified by exploiting invariance in the $xy$ plane to assume without loss of generality that the dipole in this case is aligned along the $x$ direction. Therefore we need to calculate
\begin{align}
 \Delta \Gamma_{\parallel}^\text{HS}&=2\omega_\text{A}^2|d|^2  \text{Im} \B{G}^\text{HS}_{xx}(\B{r}_\text{A},\B{r}_\text{A},\omega_\text{A})\label{GammaParallelHS},\\
\Delta  \Gamma_{\perp}^\text{HS}&= 2\omega_\text{A}^2|d|^2  \text{Im} \B{G}^\text{HS}_{zz}(\B{r}_\text{A},\B{r}_\text{A},\omega_\text{A})\label{GammaPerpHS}.
\end{align}
Using the half-space Green's function \eqref{HalfSpaceGF} we find:
\begin{align}
 \B{G}^\text{HS}_{zz}(\B{r},\B{r}',\omega_\text{A}) &= \frac{i}{4\pi} \int_0^\infty  dk_\parallel  \frac{ k_\parallel^3 }{ \omega^2  k_z} e^{i k_z(z-z')} R_\text{TM} e^{2 i k_z z'},\notag \\
 \B{G}^\text{HS}_{xx}(\B{r},\B{r}',\omega_\text{A}) &=\frac{i}{8\pi}\int_0^\infty dk_\parallel \frac{k_\parallel}{{ \omega ^2 k_z}}  e^{i k_z (z-z')}\notag \\
& \qquad  \times e^{2 i k_z z' } \left(\omega ^2 R_\text{TE}-k_z^2 R_\text{TM}\right).
\end{align}
Substituting these into eqs.~ \eqref{GammaParallelHS} and \eqref{GammaPerpHS} and evaluating the integrals in the same way as for the free space case, one eventually finds the following results the decay rates near a perfect conductor ($R_\text{TE} \to -1$, $R_\text{TM}\to 1$)
\begin{align}
\Delta  \Gamma_{\parallel}^\text{HS} &=\frac{|d|^2}{16 \pi  z^3} \Big[\left(1-4 \omega_\text{A} ^2 z^2\right) \sin (2 \omega_\text{A}  z) \notag\\
&\qquad \qquad\qquad \qquad\qquad -2 \omega_\text{A}  z \cos (2 \omega_\text{A}  z) \Big], \\
\Delta  \Gamma_{\perp}^\text{HS} &= \frac{|d|^2}{8 \pi  z^3} \left[\sin (2 \omega_\text{A}  z)-2\omega_\text{A}  z \cos (2 \omega_\text{A}  z)\right] ,
\end{align}
in agreement with \cite{Meschede:1990tg}. The $z$-dependence of $\Gamma_\parallel^\text{HS} $ and $\Gamma_\perp^\text{HS} $  is shown in Fig.~\ref{DecayRateHalfSpace}.
This shows the well-known property that an atom whose dipole moment is aligned perpendicular to a perfectly reflecting surface has its decay rate enhanced by a factor of two in the small-distance limit. Similarly, an atom whose dipole moment is aligned parallel to such a surface has its decay rate completely suppressed as it approaches the boundary. Far away from the surface the free-space value is recovered in both cases as expected. 

\begin{figure}[h!]
\includegraphics[width = \columnwidth]{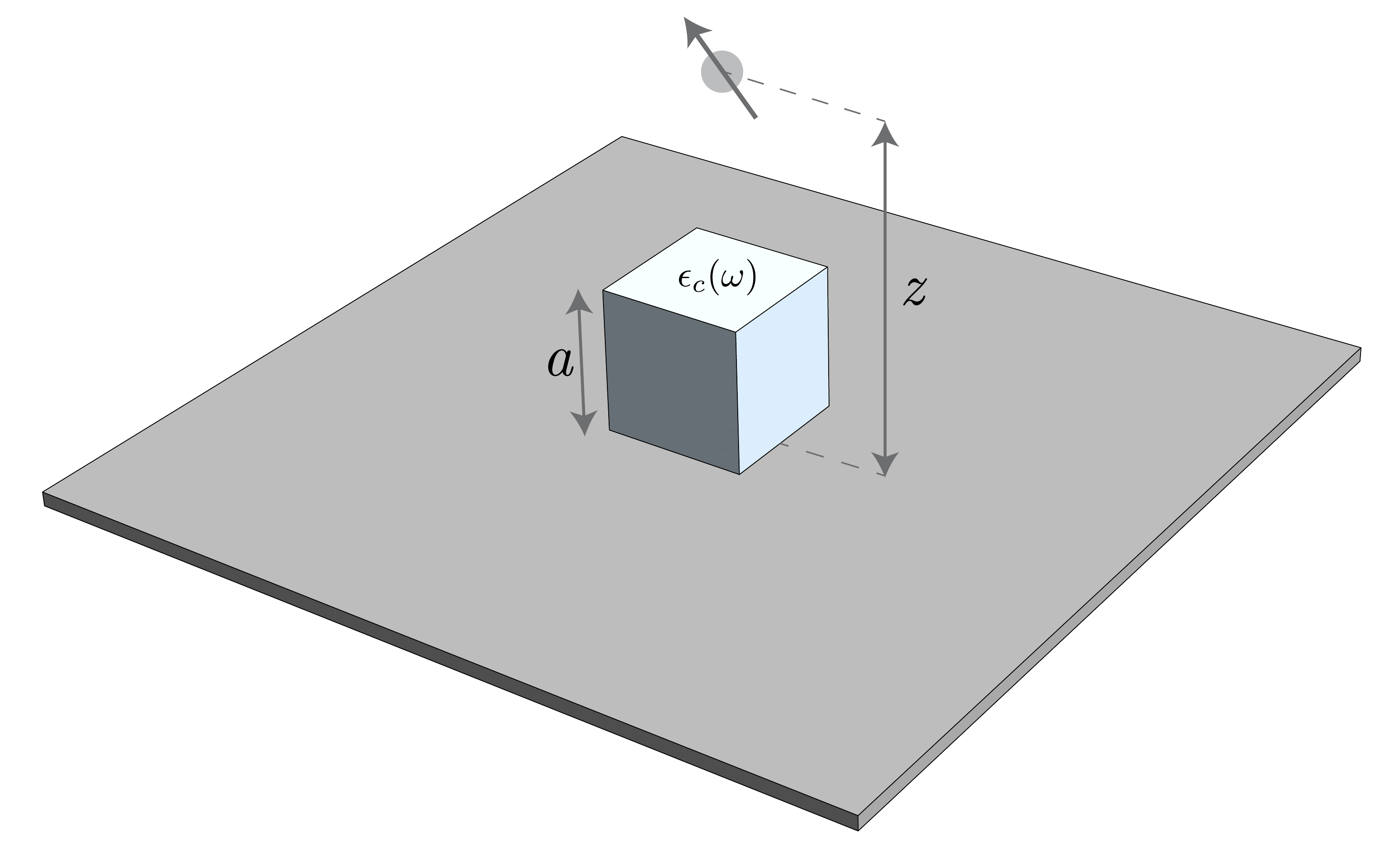}
\caption{Cubic deposition geometry. The cube and substrate can be made of different materials, the only restriction on their properties is that cube material must be weakly dielectric in order for the Born series to converge. We emphasise that the substrate can be made of any desired material since it does not take part in the perturbative approximation. } \label{CubePicture}
\end{figure}

We will now use our modified Green's function \eqref{ModGFunction} to produce new results for more complicated geometries, using the above known results as points of comparison. The new geometry that we choose is a cube of side $a$ and refractive index $\epsilon_c(\omega)$ deposited on a half-space, as shown in Figure \ref{CubePicture}.
This means the volume integral over $\B{s}$ in Eq.~\eqref{ModGFunction} becomes
\begin{equation}
 \int_V d^3 \B{s} \to \int_{-a/2}^{a/2} ds_x \int_{-a/2}^{a/2} ds_x \int_{0}^a ds_z \; . \end{equation}

Part of the reason for choosing this shape in particular is that, as mentioned in the introduction, the method presented here does not break down for geometries with sharp corners, in contrast to other approaches to radiative corrections near perturbed half-spaces which rely on the surface being smooth in some sense  \cite{Bimonte:2014fw, Messina:2009gv}.  As we will see later on, the approach used here can produce highly non-trivial results in the regions near sharp objects. 

Taking the modified Green's function \eqref{ModGFunction} and transforming to polar co-ordinates $\{k_x,k_y\} = \{k_\parallel \sin \phi, k_\parallel \cos \phi \}$ (with similar for the primed co-ordinates) we find for the $xx$ and $zz$ components of the modified Green's function in the limit where the substrate is perfectly reflecting:
\begin{align} \label{DecayRateIntegrandxx}
&\Delta  \wholeGF^{\text{cube}}_{\text{PM},xx}(\B{r},\B{r}',\omega) = \frac{\delta\epsilon_c(\omega)}{16 \pi ^4}\int_0^\infty \!\!\!dk_\parallel   \int_0^\infty \!\!\!dk_\parallel'\int_0^{2\pi} \!\!\!\!d\phi \int_0^{2\pi}\!\!\!\!d\phi'\notag \\
&\times \frac{k_\parallel  k_\parallel'}{ \omega ^2 k_z k_z' } \frac{k_z \cos \left(a k_z\right) \sin \left(a k_z'\right)-k_z' \sin \left(a k_z\right) \cos \left(a k_z'\right)}{(k_\parallel^2-k_\parallel'^2 ) (\cosAbbrv k_\parallel-\cosAbbrv' k_\parallel') (\sinAbbrv k_\parallel-\sinAbbrv' k'_\parallel) }   \notag
\\
&\times \left[\left(\cosAbbrv^2-\sinAbbrv^2-1\right) k_\parallel^2+2 \omega ^2\right] \sin \left[\nicefrac{a}{2}  \left(\cosAbbrv k_\parallel-\cosAbbrv ' k'_\parallel\right)\right]\notag \\
& \times \left[\left(\cosAbbrv'^2-\sinAbbrv'^2-1\right) k_\parallel'^2+2 \omega ^2\right] \sin \left[\nicefrac{a}{2}  \left(\sinAbbrv k_\parallel-\sinAbbrv ' k'_\parallel\right)\right] \notag\\  
&\times e^{i \left[r_x ( \sinAbbrv k_\parallel- \sinAbbrv' k'_\parallel)+r_y(\cosAbbrv k_\parallel - \cosAbbrv' k'_\parallel)+k_z' r_z'+k_z r_z\right]},
\end{align}
and
\begin{align} \label{DecayRateIntegrandzz}
&\Delta  \wholeGF^{\text{cube}}_{\text{PM},zz}(\B{r},\B{r}',\omega) = \frac{\delta\epsilon_c(\omega)}{4 \pi ^4}\int_0^\infty \!\!\!dk_\parallel   \int_0^\infty \!\!\!dk_\parallel'\int_0^{2\pi} \!\!\!\!d\phi \int_0^{2\pi}\!\!\!\!d\phi'\notag \\
& \times \frac{k_\parallel^3 k'^3_\parallel}{\omega ^2 k_z k_z' }\frac{ k_z \sin \left(a k_z\right) \cos (a k_z')-k_z' \cos (a k_z) \sin (a k_z') }{ (k_\parallel^2-k'^2_\parallel) (\cosAbbrv k_\parallel-\cosAbbrv' k'_\parallel) (\sinAbbrv k_\parallel-\sinAbbrv' k'_\parallel)}\notag \\
& \times \sin \left[\nicefrac{a}{2} \left(\cosAbbrv k_\parallel-\cosAbbrv' k'_\parallel\right)\right] \sin \left[\nicefrac{a}{2} \left(\sinAbbrv k_\parallel-\sinAbbrv' k'_\parallel\right)\right]\notag \\
&\times e^{i \left[r_x ( \sinAbbrv k_\parallel- \sinAbbrv' k'_\parallel)+r_y(\cosAbbrv k_\parallel - \cosAbbrv' k'_\parallel)+k_z' r_z'+k_z r_z\right]},
\end{align}
where we have abbreviated
\begin{align}
&\cosAbbrv \equiv \cos \phi, &  \cosAbbrv' &\equiv \cos \phi' \notag ,  \\
&\sinAbbrv \equiv \sin \phi, &  \sinAbbrv' &\equiv \sin \phi',
\end{align}
and immediately taken the parallel coincidence limit $\B{r}_\parallel' \to \B{r}_\parallel$. The perturbative approximation holds as long as $\epsilon_c(\omega_\text{A})-1<1$, meaning that we require $\epsilon_c(\omega_\text{A})<2$. In practice this means that the absorption lines of the medium constituting the cube must be well-separated in frequency from the relevant atomic transition frequency $\omega_\text{A}$. Here we will simply choose $\epsilon_c(\omega_\text{A})$ to be such that the condition $\epsilon_c(\omega_\text{A})<2$ holds. 

\begin{figure}[h!]
\includegraphics[width = \columnwidth]{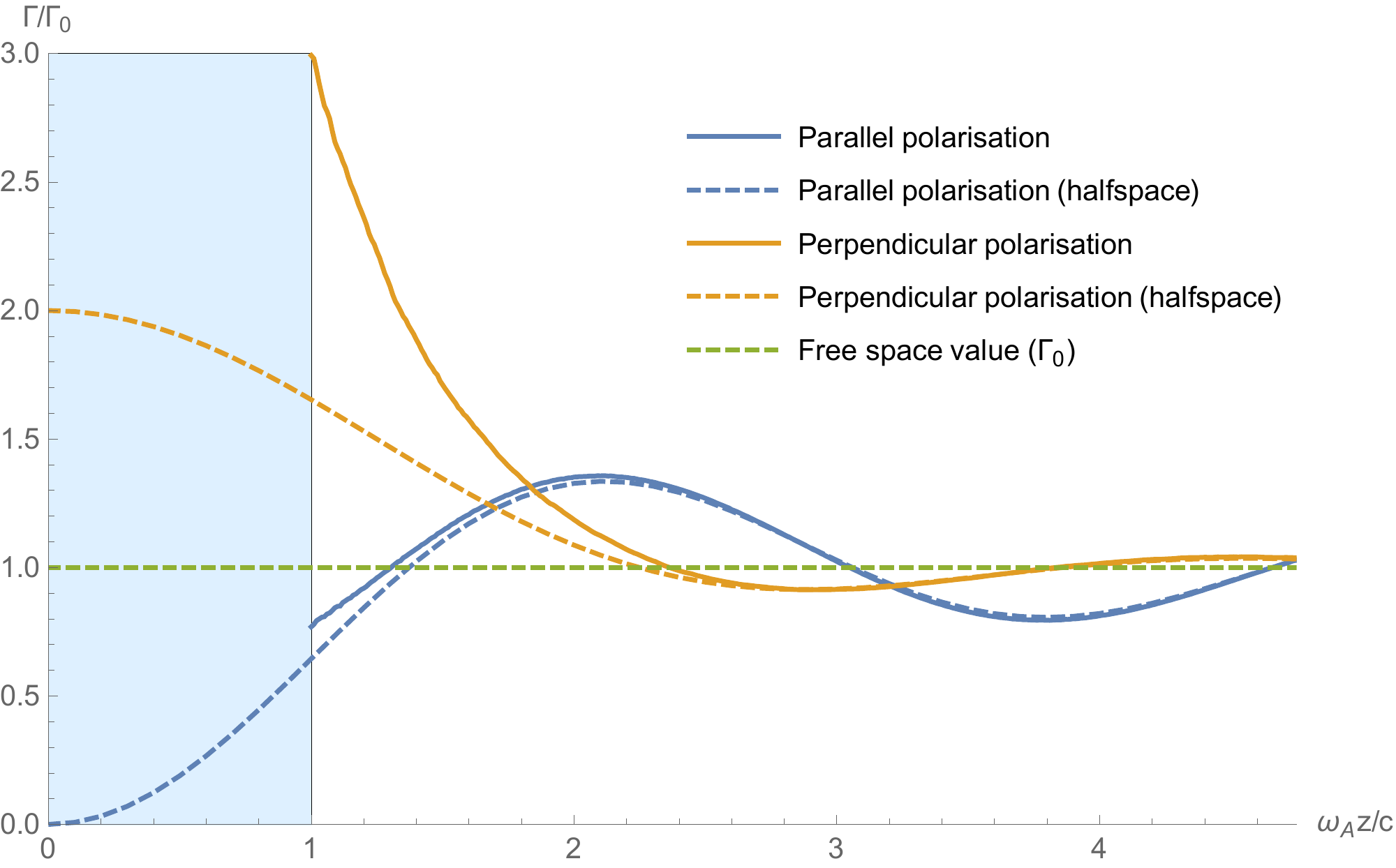}
\caption{Modified decay rates (solid lines) near a cube of dielectric constant $\epsilon_c(\omega_A) =1.8$ deposited on a perfectly reflecting half-space, with the decay rates for the bare halfspace and free space (dashed lines) shown for comparison .  The shaded area represents the depth of the cube added to the halfspace. We do not present results for the region interior to the cube because local-field effects \cite{Knoester:1989kk, Barnett:1992gf, Scheel:1999eq} would come into play there, but these are beyond the scope of this work.}  \label{DecayRatesCubeVaryZ}
\end{figure}

\begin{figure}[h!]
\includegraphics[width = \columnwidth]{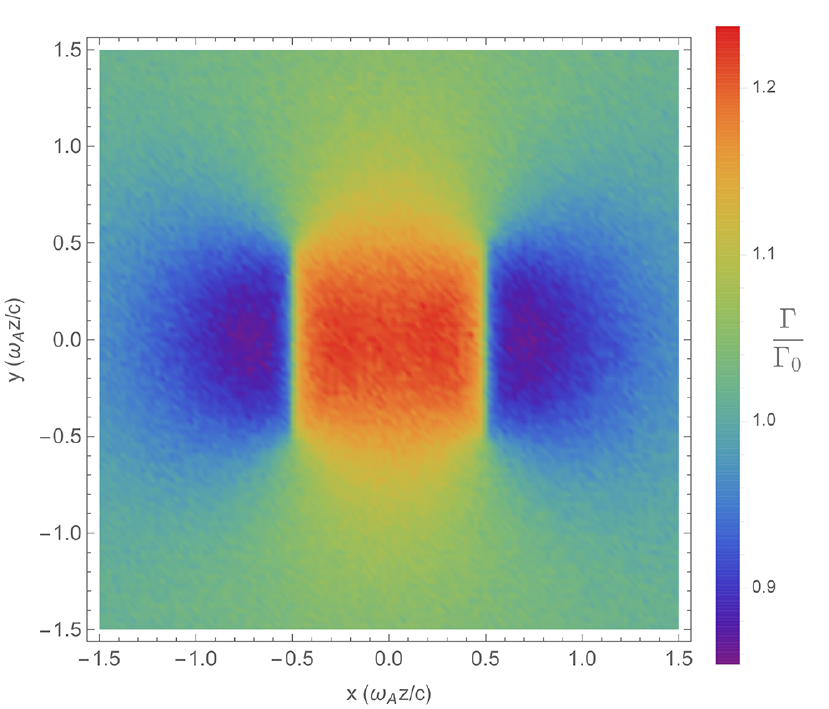}
\caption{Normalised decay rate for an $x$-polarised atom at a distance $0.01a$ above a cube of side length $a=1$ (in dimensionless units $\omega_\text{A} z/c$) and refractive index $\epsilon_c(\omega) =1.8$ deposited on a perfectly reflecting half-space. The decay rate is expressed in units of the decay rate at the same distance above the `bare' perfectly reflecting half-space (i.e. that with no cubic deposition).} \label{XYDecay}
\end{figure}

The quadruple integrals \eqref{DecayRateIntegrandxx} and \eqref{DecayRateIntegrandzz} are straightforward to numerically evaluate in ready-made software such as Mathematica or Maple --- no specialised numerical techniques are required. Their ease of evaluation  arises because the angular integrals are over a finite range and the $k_\parallel$ integrals are exponentially damped at infinity. We present a selection of results of this numerical study in Figs.~\ref{DecayRatesCubeVaryZ} and \ref{XYDecay}. 
We note in particular that Fig.~\ref{XYDecay} shows the highly non trivial position-dependence of the decay rate --- for example the decay rate can be enhanced or suppressed (relative to the value near a bare halfspace) depending on the precise position of the atom in the plane above the cube.

\section{Casimir-Polder potential of a finite grating}
\subsection{Background and motivation}
We now turn our attention to a more complex but experimentally-relevant situation, namely the Casimir-Polder (CP) potential of an atom near a surface, as first described in \cite{Casimir:1948bd}. The CP potential results from the modification of the level structure of a polarizable atom by a surface-dependent quantised field ---  it is the surface-dependent version of the Lamb shift. The resultant force has been measured to high precision \cite{Sukenik:1993ir} and is of increasing importance in emerging quantum technologies \cite{Judd:2011kq}. The calculation is inherently more complicated than that for the decay rate in section \ref{DecayRateSection}. As we shall se, this is largely because the potential depends on a sum over all photon frequencies, rather than being determined by a specific transition frequency like the decay rate. An additional complication is that calculation of a CP potential involves subtraction of the contribution of the homogenous part of the Green's function at each particular point in order to extract a geometry-dependence. This is necessary because, unlike the decay rate, evaluation of a CP potential in free space (i.e., the Lamb shift) requires a completely different full field-theoretic approach. As detailed in the introduction, care must be taken with CP potentials in this Born-series approach because of the interplay between this subtraction of a homogenous part and the perturbative approximation. 

\begin{figure}[h!]
\includegraphics[width =0.9 \columnwidth]{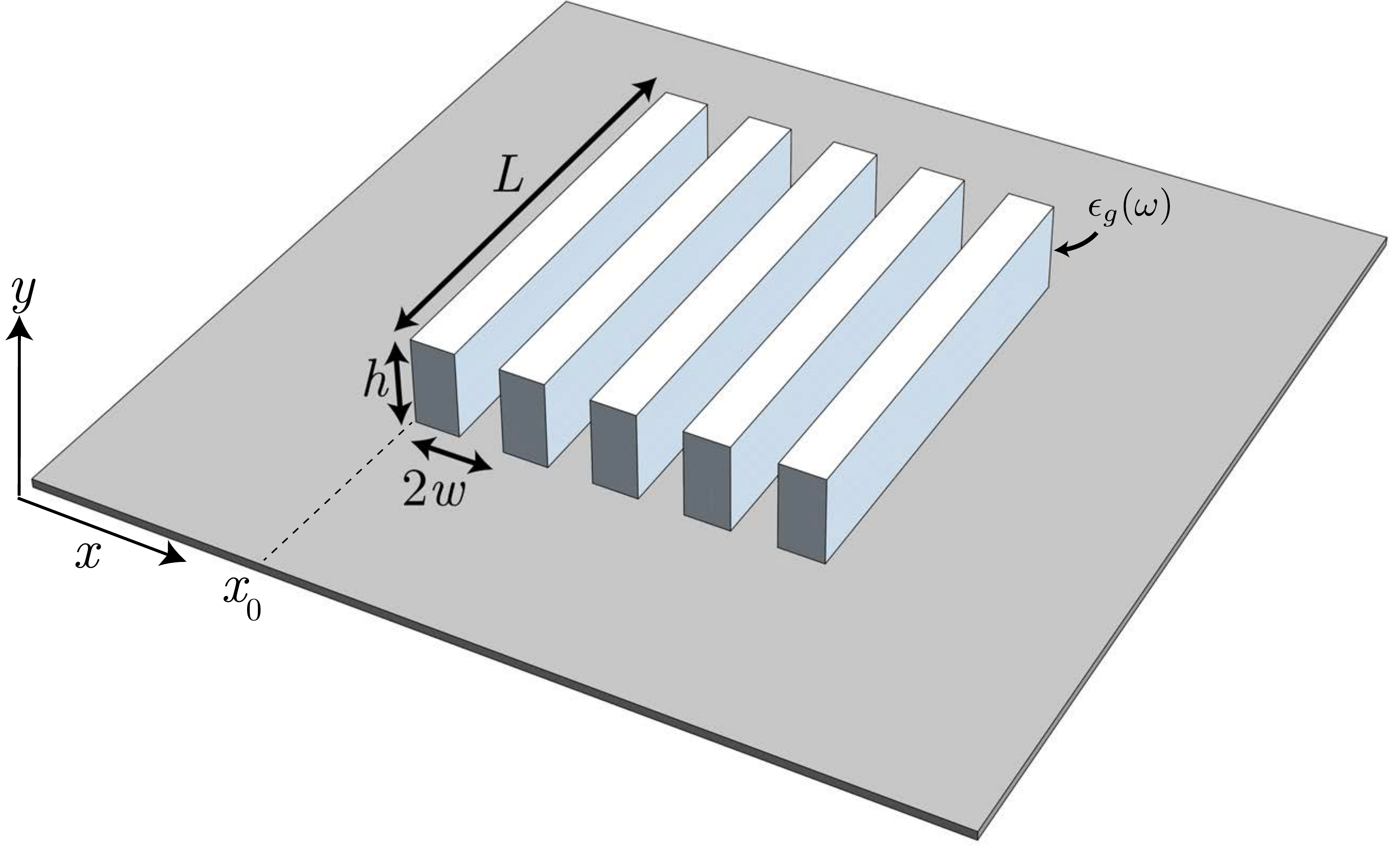}
\caption{Geometry of the finite grating considered here. Just as for the cubic deposition shown in fig.~\ref{CubePicture}, the substrate can be made of any desired material, again the only restriction on the system  is that the grating material must be weakly dielectric. While our method works for any number of grooves $N$, we will choose $N=5$ here and throughout as shown in the figure.} \label{GratingPictureLabelled}
\end{figure}

We will calculate the CP potential in vacuum near an $N$-grooved finite grating, like that shown in Fig.~\ref{GratingPictureLabelled}. This choice is  motivated by the structures used ongoing experiments in atom optics and matter-wave interferometry such as \cite{Nshii:2013fv} and \cite{Gunther:2007jy}.
There is section of existing literature on CP forces near periodic gratings \cite{ContrerasReyes:2010bz, Lussange:2012ct}, however these works take advantage of the Bloch theorem and so are only strictly applicable to infinite, precisely periodic gratings which are not necessarily good approximations to real experiments. In fact, as we will see later, non-trivial behaviour of the CP potential occurs \emph{outside} the immediate vicinity of the grating, which of course cannot be seen if the grating is assumed to be infinite. 

\subsection{Expressions for Casimir-Polder potential}

The CP potential $U_\text{CP}$ for an isotropically polarisable atom at position $\B{r}_\text{A}=(x_\text{A},y_\text{A},z_\text{A})$ in a region with \emph{scattering} electromagnetic Green's function $\B{G}$ may be written in terms of an integral over complex frequency $\xi$ as \cite{Wylie:1984cz, Scheel:2008tu} 
\begin{equation} \label{CPPotentialBasic}
U_\text{CP}(\B{r}_\text{A}) = \frac{1}{2\pi} \int_0^\infty d\xi \, \xi^2 \alpha(i\xi) \text{Tr}\, \B{G}(\B{r}_\text{A},\B{r}_\text{A},i\xi)\; ,
\end{equation}
where $\alpha$ is the polarisability of a ground-to-excited atomic state transition of frequency $\omega_{ij}$ and dipole moment $d_{ij}$ and is given by
\begin{equation}
\alpha (\omega) = \frac{2}{3} \lim_{\varepsilon \to 0} \frac{\omega_{ij}|d_{ij}|^2}{\omega_{ij}^2 -\omega^2 -i \omega \varepsilon} \; , 
\end{equation}
where $\varepsilon$ is a real infinitesimal \footnote{The infinitesimal $\varepsilon$ should not be confused with the dielectric constant $\epsilon$} related to the line width of the atomic state \cite{Milonni:2004bz}. The Green's function is to be taken with both spatial arguments equal to the position $\B{r}_\text{A}$ of the atom, this is to be understood as a limiting value. Just as in the decay rate calculation in section \ref{DecayRateSection}, we will use a selection of standard results as points of comparison for later results. The first of these is the Casimir-polder potential at a distance $z_\text{A}$ from a perfectly conducting plane in the non-retarded regime, This regime is where the round-trip time for a photon to travel from the atom to the surface and back is much smaller than the timescale associated with the atomic frequency. In other words it is the small-distance approximation if the atomic transition frequency is assumed to be a fixed constant. The well-known result in this regime is \cite{Casimir:1948bd};
\begin{equation}\label{U0}
U_{\text{CP}0}(\omega_\text{A} z_\text{A} \ll 1) = -\frac{|d_{ij}|^2}{48\pi z_\text{A}^3} \; . 
\end{equation}
The second quantity we will use as a comparison is the force that an atom in this potential experiences, namely
\begin{align}\label{F0}
F_{\text{CP}0}(\omega_\text{A} z_\text{A} \ll 1) &=-\frac{\partial}{\partial z_\text{A}} U_{\text{CP}0}(\omega_\text{A} z_\text{A} \ll 1) \notag \\ &=  -\frac{|d_{ij}|^2}{16\pi z_\text{A}^4} \, , 
\end{align}
which a statement the well-known Casimir-Polder force of attraction between a polarizable atom and surface, in this case in the non-retarded regime and for a perfectly conducting material.

Equation \eqref{BornExp} tells us that the Green's function that encodes the behaviour of the EM field near the grating is given by the sum of two terms: $\wholeGF^{\text{HS}}$ which describes the unperturbed half-space and $\Delta\wholeGF^{\text{HS}}(\B{r},\B{r}',\omega)$ which describes the correction resulting from deposition of the grating on its surface. The CP potential \eqref{CPPotentialBasic} requires the use of a scattering Green's function. Since the whole Green's function $\wholeGF$ is linear in its two contributions $\wholeGF^{\text{HS}}$ and $\Delta\wholeGF^{\text{HS}}(\B{r},\B{r}',\omega)$ it suffices to find the scattering parts $\B{G}^{\text{HS}}$ and $\Delta \B{G}^{\text{HS}}(\B{r},\B{r}',\omega)$ of these two contributions separately, which together give the scattering Green's function $\B{G}$. Also, the linearity of the CP potential in the scattering Green's function means that we can find the contributions from the two scattering parts separately. We will confine ourselves to the most physically-relevant region $z,z'>0$, meaning that the subtraction of a homogenous part is achieved by setting all reflection coefficients in the Green's function to zero and subtracting the resulting quantity. For the unperturbed part $\B{G}^{\text{HS}}$ we have:
\begin{equation}
\B{G}^{\text{HS}}(\B{r},\B{r}',\omega)=\wholeGF^{\text{HS}}(\B{r},\B{r}',\omega)-\wholeGF^{\text{HS}}_{R_\sigma\to0}(\B{r},\B{r}',\omega)\; ,
\end{equation}
where $\wholeGF^{\text{HS}}_{R_\sigma\to0}(\B{r},\B{r}',\omega)$ coincides with the Green's function of free space since the region near the grating as assumed to be vacuum. 

The scattering part of the Green's function correction \eqref{ModGFunction} is obtained by subtracting the portion that remains when setting all reflection coefficients to zero. Consequently, isolating the scattering part $\Delta \B{G}^\text{HS}(\B{r},\B{r}',\omega)$ of $\Delta \wholeGF(\B{r},\B{r}',\omega)$ is trivial because of the way it is stated in Eq.~\eqref{ModGFunction} --- all one needs to to is remove the term independent of reflection coefficients, giving:
\begin{align} \label{DeltaGHS}
&\Delta\B{G}^{\text{HS}}_{ij}(\B{r},\B{r}',\omega) =  \int_V d^3 \B{s}  \int d^2  \B{k}_\parallel  \int d^2 \B{k}'_\parallel P\notag \\
&\times\left[K_{ij}^\text{TETM} R_\text{TE} R_\text{TM}+\sum_{\sigma} (K^\sigma_{ij} R_\sigma+K^{\sigma\sigma}_{ij} R^2_\sigma) \right]\, ,\end{align}
with matrix elements $K_{ij}$ listed in Appendix \ref{GFMatrixElements}. Now that we have the Green's function correction \eqref{DeltaGHS} we can find the correction to the CP potential resulting from the deposition of the grating on the half-space from

\begin{equation} \label{CPPotentialDelta}
\Delta U_\text{CP}(\B{r}_\text{A}) = \frac{1}{2\pi} \int_0^\infty d\xi \, \xi^2 \alpha(i\xi) \text{Tr}\, \Delta \B{G}(\B{r}_\text{A},\B{r}_\text{A},i\xi)\; ,
\end{equation}

\subsection{Grating results and discussion}

The volume integral describing the $N$-grooved grating shown in Fig.~\ref{GratingPictureLabelled} is:
\begin{equation} \label{VolIntegralGrating}
 \int_V d^3 \B{s} \to \sum_{n=0}^{N-1}  \int_{x_0+2nw}^{x_0+(2n+1)w} ds_x \int_{-L/2}^{L/2} ds_y \int_0^h ds_z \; ,
\end{equation}
with $x_0=-w(N -\nicefrac{3}{4})$ if the grating is such that the centre of the base of the middle groove is at $s_x=0$.

\begin{figure}
\includegraphics[width = \columnwidth]{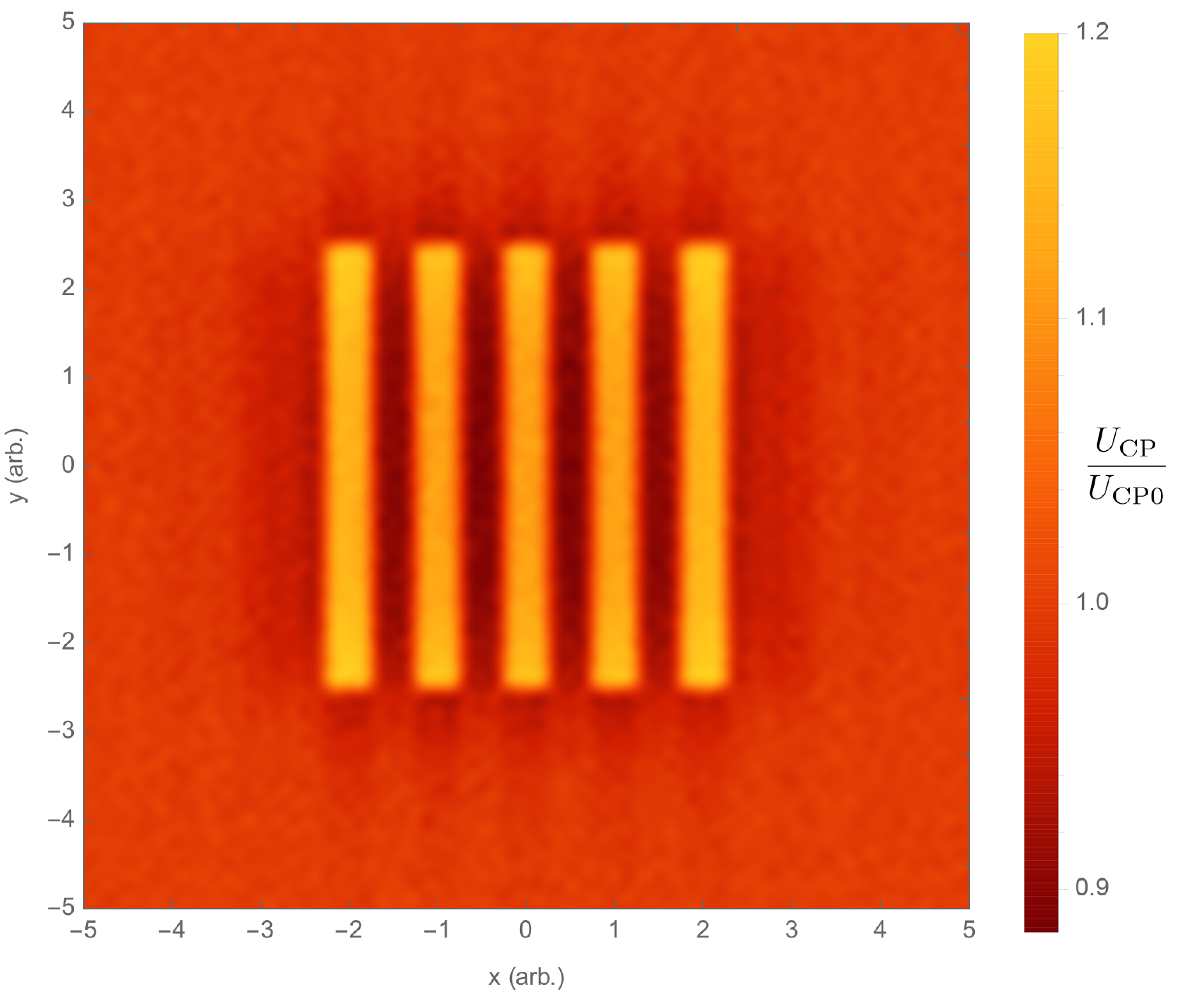}
\caption{Non-retarded Casimir-Polder potential above the grating shown in Fig.~\ref{GratingPictureLabelled}. The potential is shown in units of the perfect reflector potential \eqref{U0}, which is the potential if the grating had not been deposited on the half-space. All length scales in the problem can be expressed in terms of a reference length which cancels out (in the non-retarded regime) when normalizing to the bare perfect reflector result. Thus, the units on both axes of the above plot are arbitrary --- in other words the result remains valid whatever unit is assigned to the $x$ and $y$ axes as long as the non-retarded approximation holds. The parameters describing the grating are $h=1 , w=1 $ and $L=5$ in the same units as the $x$ and $y$ axes. The dielectric constant is $\epsilon_g = 1.8$. Almost invisible in this figure  is a suggestion of interesting behavior `outside' the grating along the $x$ axis, this is shown in detail in Fig.~\ref{ForceAndPotential}. } \label{GratingResult}
\end{figure}

\begin{figure}
\includegraphics[width = \columnwidth]{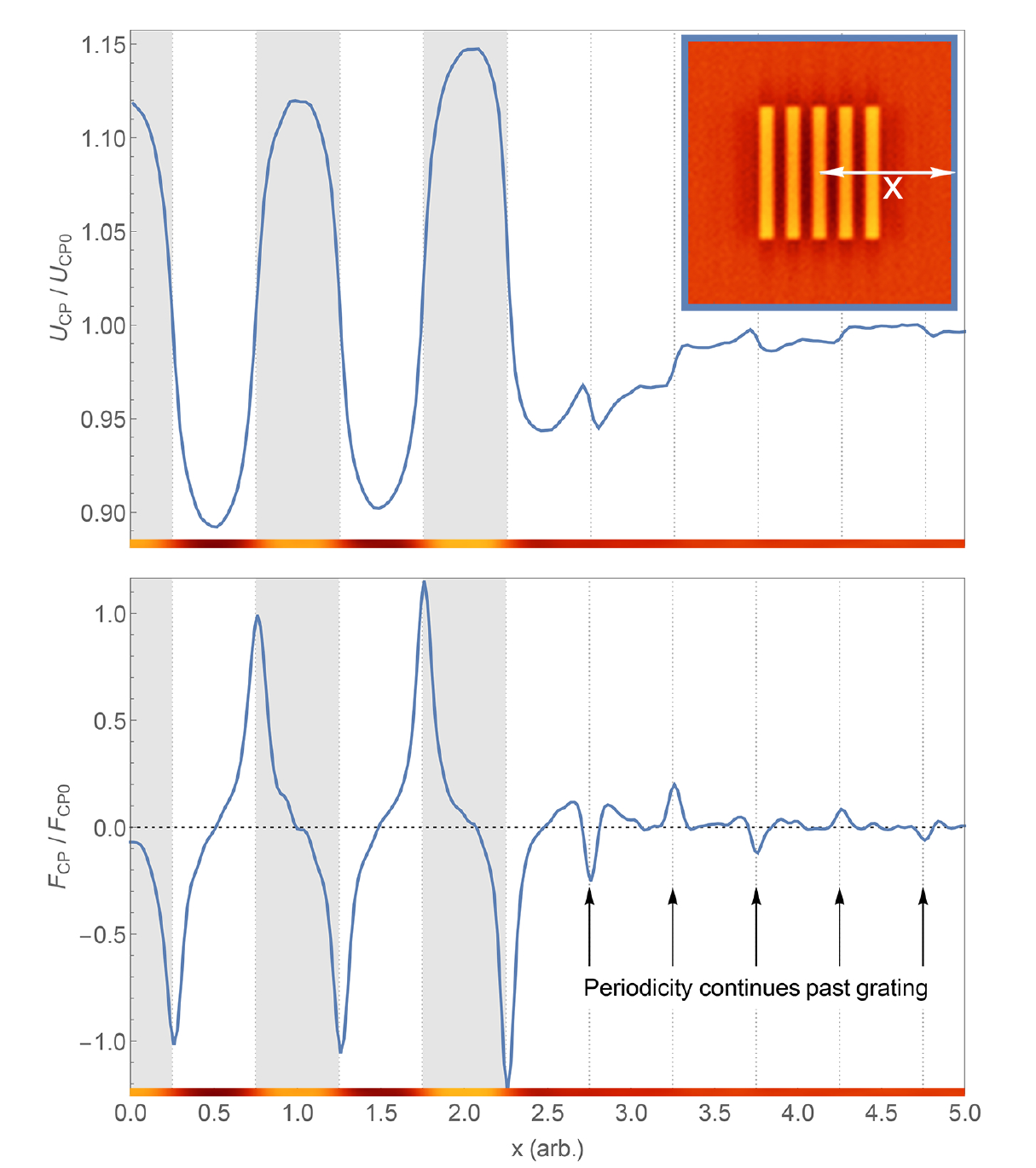}
\caption{Upper plot: Potential $U_\text{CP}$ near the grating shown in Fig.~\ref{GratingPictureLabelled}. As in Fig.~\ref{GratingResult} the potential is plotted in units of its value $U_{\text{CP0}}$ [Eq.~\eqref{U0}] near a simple planar perfect reflector, and all the parameters are the same as for Fig.~\ref{GratingResult} which is reproduced as an inset and schematically on the horizontal axes. Lower plot: Lateral ($x$-directed) force $F_\text{CP}$ near the grating shown in Fig.~\ref{GratingPictureLabelled} for the same parameters used in Fig.~\ref{GratingResult}. In the same spirit as the other plots, we normalise to what the force would have been if the grating were not present, but the lateral force without the grating would of course be zero. For this reason we use the (constant) perpendicular force \eqref{F0} as a unit instead. } \label{ForceAndPotential}
\end{figure}

For simplicity the half-space will be taken as perfectly reflecting, and the grating as non-dispersive $\epsilon_g(\omega) = \epsilon_g$ with $\epsilon_g-1<1$. We choose $N=5$, which corresponds to the grating shown in \ref{GratingPictureLabelled}. The atom's polarisability is taken to be isotropic. Using the volume elements \eqref{VolIntegralGrating} the integrals over $\B{s}$ in \eqref{CPPotentialDelta} become elementary. This leaves integrals over $k_\parallel, k_\parallel', \xi, \theta$ and $\theta'$ which may be evaluated numerically. Just as for the decay rate calculation in section \ref{DecayRateSection},  the integration is significantly simplified by the fact that the integrals over $\theta$ and $\theta'$ are both over the finite region $0..\pi/2$, and the remaining integrals are all exponentially damped. A selection of results are shown in Figs.~\ref{GratingResult} and \ref{ForceAndPotential}
The results for the CP potential directly above the grating show qualitative agreement with the infinite grating considered in \cite{ContrerasReyes:2010bz}, where it was observed that the potential is reduced between the grooves and enhanced above them, as compared to the planar result. However our results are not directly quantitatively comparable with \cite{ContrerasReyes:2010bz} due to the choices of materials made there not being consistent with our perturbative expansion. Our results for the region `outside' the grating were of course not seen even qualitatively in \cite{ContrerasReyes:2010bz} due to that work's assumption of an infinite grating. Here we have relaxed this assumption, and found that the periodicity of the Casimir-Polder potential continues laterally past the end of the grating, which to our knowledge is a previously-unseen phenomenon. 
\section{Conclusions}

In this paper we have considered some aspects of quantum electrodynamics near a surface with arbitrarily shaped features deposited on it.  The main general result is the Green's function \eqref{ModGFunction} for the perturbed half-space, which was calculated perturbatively using a Born-series expansion. We then investigated the decay rate of an atom near a cube deposited on a half-space, finding the rich position-dependence shown in Fig. \ref{XYDecay}. Finally we presented the Casimir-Polder potential of a finite grating deposited on a substrate and demonstrated the previously-unseen quality that the lateral periodicity of the potential can continue beyond the grating, as shown in Fig. \ref{ForceAndPotential}. The Green's function \eqref{ModGFunction} can be used to calculate quantum electrodynamical quantities near a half-space with \emph{any} small deposition on it, so the work presented here should have applications in a variety of ongoing and planned experiments \cite{Gunther:2007jy, Judd:2011kq, Nshii:2013fv}. 

\section{Acknowledgements}

It is a pleasure to thank the UK Engineering and Physical Sciences Research Council (EPSRC) for financial support. 

\appendix
\section{Green's function matrix elements}\label{GFMatrixElements}

There are five terms in Eq.~\eqref{ModGFunction}, each of which is a one of two $3\times3$ matrices (one for each of choice of $r_z \lessgtr s_z$), giving a total of 90 matrix elements that we, in principle, need to calculate. However there are various constraints that reduce this number significantly. Firstly, some matrix elements are not independent due to the $xy$ symmetry of the half-space. In particular;
\begin{align}
K^\tau_{yy\lessgtr} &= K^\tau_{xx\lessgtr}(k_x \leftrightarrow k_y),\notag\\
K^\tau_{yz\lessgtr} &= K^\tau_{xz\lessgtr}(k_x \leftrightarrow k_y),\notag\\
K^\tau_{zy\lessgtr} &= K^\tau_{zx\lessgtr}(k_x \leftrightarrow k_y),\notag\\
K^\tau_{yx\lessgtr} &= K^\tau_{xy\lessgtr},
\end{align}
where $\tau = (\text{TE}, \text{ TM}, \text{ TETE}, \text{ TMTM}, \text{ TETM})$. This restriction reduces the number of required matrix elements by 
\begin{equation*}
(4 \text{ constraints }) \times (5 \text{ different } \tau)\times(2 \text{ for } r_z \lessgtr s_z)=40,
\end{equation*}
leaving a total of $50$. This number can be further reduced by noting that the definition of TE modes is that they have no electric field in the $z$ direction, which ultimately means that any matrix element for TE polarisation where at least one index is $z$ is in fact identically zero
\begin{align}
K^\text{TE}_{xz\lessgtr}=K^\text{TE}_{zx\lessgtr}=K^\text{TE}_{zz\lessgtr}  &= 0 \notag, \\
 K^\text{TETE}_{xz\lessgtr} = K^\text{TETE}_{zx\lessgtr} = K^\text{TETE}_{zz\lessgtr} &= 0 \; , 
\end{align}
and similarly
\begin{equation}
K^\text{TETM}_{xz\lessgtr} = K^\text{TETM}_{zx\lessgtr} = K^\text{TETM}_{zz\lessgtr} = 0 \; 
\end{equation}
which together reduce the required number by $18$ leaving $90-40-18 = 32$ matrix elements to calculate, which can be partitioned into two groups of $16$, where each group corresponds to one choice of $r_z \lessgtr s_z$. We now simply list these matrix elements, which are obtained by application of the differential operators \eqref{DOps} to the functions $F^\sigma(z,z')$ given by Eq.~\eqref{FDefs}. For $r_z>s_z$ the matrix elements representing coefficients of terms linear in the reflection coefficients are:
\begin{align}\label{LMatrixEls}
K^\text{TE}_{xx>} &=e^{-2 i k_z s_z}\omega ^2k_y'^2\left(k_x^2 k_z^2+k_y^2 \omega ^2\right) + \text{primed}, \notag \\
K^\text{TE}_{xy>} &=e^{-2 i k_z s_z}k_x k'_x k_y k_y' \omega ^2 k_\parallel^2 + \text{primed},\notag\\
K^\text{TM}_{xx>} & =- e^{-2 i k_z s_z}k_x'^2 k_z'^2  \left(k_x^2 k_z^2+k_y^2 \omega ^2\right)+ \text{primed},\notag\\
K^\text{TM}_{xy>} & =e^{-2 i k_z s_z}  k_\parallel^2 k_x' k_y k_y' k_x k_z^{'2} + \text{primed},\notag \\
K^\text{TM}_{xz>} &=e^{-2 i k_z s_z} k_x k_x' k_z k_z' k_\parallel^2k_\parallel'^2- \text{primed} ,\notag\\
K^\text{TM}_{zx>} &= -e^{-2 i k_z s_z}k_x k_x' k_z k_z' k_\parallel^2k_\parallel'^2- \text{primed},\notag\\
K^\text{TM}_{zz>} &= e^{-2 i k_z s_z}k_\parallel^4 k_\parallel'^4+ \text{primed}
\end{align}
where `primed' is a shorthand for the quantity that precedes it with $\B{k} \to \B{k}'$ and $\B{r}\to \B{r}'$, with the latter replacement only being relevant for $r_z<s_z$ as we shall see. Continuing, the $r_z>s_z$ matrix elements representing coefficients of terms quadratic in particular reflection coefficients are
\begin{align}\label{QMatrixEls}
K^\text{TETE}_{xx>} &= k_y^2 k_y'^2 \omega^4, & K^\text{TMTM}_{xx>} &=k_x^2 k_x'^2 k_z^2 k_z'^2,\notag \\
K^\text{TETE}_{xy>}&=k_x'k_xk_y'k_y\omega^4, & K^\text{TMTM}_{xy>}&=k_xk_x'k_yk_y'k_z^2k_z'^2,\notag \\
K^\text{TMTM}_{xz>} &=k_x k_x' k_z k_z' k_\parallel^2 k_\parallel'^2, & K^\text{TMTM}_{zx>}&= k_x k_x' k_z k_z' k_\parallel^2 k_\parallel'^2,\notag\\
K^\text{TMTM}_{zz>}&= k_\parallel^4 k_\parallel'^4,
\end{align}
and finally the coefficients of the terms that mix TE and TM reflection coefficients
\begin{align} \label{MixedMatrixEls}
K^\text{TETM}_{xx>} &= -\omega ^2 \left(k_x^2 k_y'^2 k_z^2+k_x'^2 k_y^2 k_z'^2\right), \notag\\
K^\text{TETM}_{xy>} &= k_x k'_x k_y k'_y \omega ^2 \left(k_z^2+k_z'^2\right) .
\end{align}
For $r_z<s_z$ the entire set of $16$ coefficients can be obtained from Eqs.~\eqref{LMatrixEls}-\eqref{MixedMatrixEls} by taking $s_z \to r_z$ (before adding the `primed' parts), so that for example:
\begin{equation}
K^\text{TM}_{zz<} = (e^{-2 i k_z r_z}+e^{-2 i k_z r_z'}) k_\parallel^4 k_\parallel'^4.
\end{equation}
We have now completely specified all terms in $\eqref{ModGFunction}$.

%
%
%
%

%

\end{document}